\documentclass[reprint, amsmath, amssymb, aps, prapplied]{revtex4-2} 

\usepackage{graphicx} 
\usepackage{dcolumn} 
\usepackage{booktabs} 
\usepackage{multirow}
\usepackage{bm} 
\usepackage{natbib} 
\usepackage{hyperref} 
\hypersetup{
    colorlinks=true,    
    linkcolor=blue,     
    citecolor=blue,     
    filecolor=blue,     
    urlcolor=blue       
}

\begin{document}

\preprint{APS/other condensed matter}

\title{Efficient Band Structure Calculation for Transitional-Metal Dichalcogenides Using the Semiempirical Pseudopotential Method}

\author{Raj Kumar Paudel\textit{$^{1}$}}
\email{rajupdl6@gate.sinica.edu.tw}
\author{Chung-Yuan Ren\textit{$^{3}$}}%
\author{Yia-Chung Chang\textit{$^{1,2}$}}
 \email{yiachang@gate.sinica.edu.tw}
\affiliation{%
Research Center for Applied Sciences, Academia Sinica, Taipei 11529, Taiwan}%
\affiliation{
 Department of Physics, National Cheng-Kung University, Tainan 701, Taiwan
}%
\affiliation{
Department of Physics, National Kaohsiung Normal University, Kaohsiung 824, Taiwan
}%

\date{\today}

\begin{abstract}
The Semiempirical Pseudopotential Method (SEPM) has emerged as a valuable tool for accurately determining band structures, especially in the realm of low-dimensional materials. SEPM operates by utilizing atomic pseudopotentials, which are derived from DFT calculations. SEPM calculations offer a unique advantage compared to DFT as they eliminate the requirement for iterative self-consistent solutions in solving the Schrödinger equation, leading to a substantial reduction in computational complexity. The incorporation of both non-local and local Semiempirical Pseudopotentials in our current approach yields band structures and wavefunctions with enhanced precision compared to traditional empirical methods. When applied to monolayer TMDCs, adjusting the parameters to align with pertinent values obtained from DFT computations enables us to faithfully replicate the band structure, opening avenues for investigating the optoelectronic properties of TMDCs and exploring their potential applications in nanodevices.
\end{abstract}

 \maketitle

\section{INTRODUCTION }
Layered transition-metal dichalcogenides (TMDCs) are a class of two-dimensional (2D) materials where covalently bound layers are stacked together by van der Waals forces, fundamentally different from their bulk counterparts. The discovery of graphene in 2004 \cite{novoselov2004electric} ignited extensive research into 2D materials due to its exceptional properties, but its lack of a band gap limits its use in logical circuits. This has shifted focus to layer TMDCs \cite{mak2010atomically,wang2012electronics,mattheiss1973band} such as
MoS$_2$, MoSe$_2$, MoTe$_2$, WS$_2$, WSe$_2$, and WTe$_2$, which offer diverse electronic properties, ranging from semiconducting states to superconductors \cite{qi2016superconductivity}. With direct band gaps for monolayers \cite{mak2010atomically}, TMDCs are ideal for next-generation  optoelectronics \cite{liu2014optical,wang2012electronics,qiu2013optical}, mechanical \cite{johari2012tuning,shi2018mechanical},sprintronics \cite{zhu2011giant,yun2022escalating} and electrocatalysis \cite{li2018metallic},  systems for energy storage and conversion. These materials also exhibit unique topological properties \cite{wu2019topological}, making them promising for novel electronic and quantum computing applications. Recent advances include the successful fabrication of high-performance field-effect transistors (FETs) \cite{chen2020environmental}, photo transistors \cite{li2014metal}, and gas sensors \cite{ko2016improvement}, highlighting the potential of TMDCs in advanced technological applications and attracting significant attention from both scientific and industrial communities.

For electronic-structure calculations, first-principles methods based on density functional theory (DFT) \cite{hohenberg1964inhomogeneous,kohn1965self,martin2020electronic,cohen2016fundamentals,ren2022density,ren2023density} and the pseudopotential (PP) scheme \cite{herring1940new,phillips1959new,chelikowsky1976nonlocal,wang1995local,wang1996pseudopotential,bester2008electronic} are widely used. However, for 2D materials like TMDCs, conventional approaches using three-dimensional plane waves (PWs) face challenges due to the need for large vacuum spaces in super-cell simulations to mimic 2D periodicity. Li et al. \cite{li1994electronic,chang1996planar} proposed a planar-basis approach that combines plane waves in the periodic directions (x-y plane) with Gaussian functions in the non-periodic direction (z), allowing for accurate calculations of total energy and work function for isolated slabs. Ren et al. \cite{ren2015mixed,ren2022density} further refined this method with a mixed-basis approach using B-spline functions \cite{deBoor1987practical} along the z-direction, offering advantages in preserving local layer-like geometry and reducing computational overhead in Kohn–Sham Hamiltonian diagonalization for charged systems.

The primary motivation for developing the Semiempirical Pseudopotential Method (SEPM) \cite{wang1996pseudopotential,paudel2023semi} lies in the computational efficiency achieved by avoiding the self-consistent density optimization central to Density-Functional Theory (DFT). This approach is particularly beneficial for nanoscale structures comprising thousands to hundreds of thousands of atoms. The concept of replacing the strong Coulomb potential with a weaker pseudopotential is well-established. Initially, empirical pseudopotentials  \cite{cohen1966band,chelikowsky1976nonlocal,pandey1974nonlocal} were fitted to match experimentally determined energy levels, allowing accurate predictions of band structures and optical properties with minimal Fourier components. However, these pseudopotentials lacked transferability across different structures. This issue was addressed by developing continuous pseudopotentials dependent on G vector lengths, enabling accurate calculations of electronic, optical, and transport properties in nanostructures using advanced computational methods \cite{wang1995local,fu1997local,bester2008electronic,molina2012semiempirical}.

 SEPM not only bypasses the need for a self-consistent solution but also allows focusing on specific parts of the eigenvalue spectrum, making the number of bands independent of the system size. This characteristic is ideal for studying optical properties or transport, where only the energy states around the band gap are involved. Additionally, by incorporating both local and non-local pseudopotentials, the Semimpirical Pseudopotential Method (SEPM) enhances the precision of band structures and wavefunctions.

In our prior research \cite{paudel2023semi}, SEPM was introduced for 2D systems such as graphene and armchair graphene nanoribbons (aGNRs) within a mixed basis approach. With just a few parameters, we successfully replicated the complete band structure of graphene and aGNRs obtained through DFT, with minimal deviation. In this study, we aim to build upon our previous work by extending the implementation of SEPM to monolayer and bilayer transition metal dichalcogenides, specifically MoS$_2$, WS$_2$, MoSe$_2$, and WSe$_2$.

The rest of this paper is organized as follows. Section II outlines the formalism of the SEPM method, detailing the construction of the SEPM pseudopotential and defining all relevant quantities used later in figures and tables. Section III applies the method to layered TMDC materials, analyzing various approximation steps and comparing numerous electronic properties calculated by our method with those obtained from well-established DFT methods, presenting the numerical results in tables and figures. Section IV focuses on calculating and comparing the electronic band structure of these materials using our method versus ab initio calculations, discussing the potential applications of SEPM. Finally, Section V concludes the paper.
\section{METHODOLOGY}
\subsection{\label{sec:level2}B-Spline Basis}
B-splines (BSs) basis \cite{deBoor1987practical,bachau2001applications} are advantageous for accurately represent both discrete and continuous electronic states, outperforming conventional methods like Slater Type Orbitals (STOs) or Gaussian Type Orbitals (GTOs). Their effectiveness lies in their ability to capture the oscillatory behavior of electronic wave-functions over large distances using localized piecewise polynomials. BSs serve as basis functions for expanding Kohn-Sham eigenstates perpendicular to the surface (along z-direction) calculations. BSs serve as basis functions for expanding Kohn-Sham eigenstates perpendicular to the surface (along the z-direction). They are easy to evaluate, including their derivatives, and offer flexibility through adjustable breakpoints, which is useful for representing rapidly changing wavefunctions. Unlike Gaussian functions or atomic orbitals fixed at specific atomic positions, BSs are position-independent, simplifying atomic structure optimizations.

B-splines are one-variable piecewise polynomial functions characterized by their polynomial order \(k\) (maximum degree \(k-1\)), designed to approximate arbitrary functions on some finite interval \([0, R_{\text{max}}]\). The B-spline basis functions are described by the following set of quantities:

\begin{enumerate}
    \item Radial Interval Division: The interval \([0, R_{\text{max}}]\) is divided into \(l\) adjacent subintervals \(I_j = [\xi_j, \xi_{j+1}]\) by a sequence of \(l + 1\) points \( \xi_j \) in strict ascending order \(0 = \xi_1 < \xi_2 < \ldots < \xi_{l+1} = R_{\text{max}}\). These points \( \xi_j \), also referred to as breakpoints (bps), are associated with a second sequence of non-negative integers \( \nu_j \) (\(j = 2, \ldots, l\)), defining continuity conditions \( C_{\nu_j-1} \) at each breakpoint.
    
    \item Knot Sequence: Another sequence of points \( \{t_i\} \) in ascending order, called knots, is used to construct the B-spline basis set of order \(k\). The B-spline basis functions are recursively defined using the knot sequence according to:
\begin{equation}
\begin{split}
B_{i\kappa}(z) &= \left( \frac{x - t_{i}}{t_{i + \kappa - 1} - t_{i}} \right)B_{i,\kappa - 1}(z) \\
&+ \left( \frac{t_{i + \kappa} - z}{t_{i + k} - t_{i + 1}} \right)B_{i + 1,\kappa - 1}(z)
\end{split}
\end{equation}

with
\begin{equation}
B_{i1}(z) = \begin{cases}
1 & \text{if } t_{i} \leq z < t_{i + \kappa} \\
0 & \text{otherwise}
\end{cases},
\end{equation}
where $i = 1,2,3,\ldots$ up to the number of knot sequences.
\end{enumerate}
The first derivative of the B-spline of order \(\kappa\) is given by
\begin{equation}
\begin{aligned}
\frac{d}{dz}B_{i\kappa}(z) = 
& \left( \frac{\kappa - 1}{t_{i + \kappa - 1} - t_{i}} \right)B_{i,\kappa - 1}(z) \\
& - \left( \frac{\kappa - 1}{t_{i + \kappa} - t_{i + 1}} \right)B_{i + 1,\kappa - 1}(z).
\end{aligned}
\end{equation}
The derivative of B-splines of order $\kappa$ is a linear combination of B-splines of order $\kappa - 1$, which is also a simple polynomial and is continuous across the knot sequence. Here, we adopt $\kappa = 4$.
B-spline functions are evaluated recursively, starting from lower-degree functions and progressing up to degree $\kappa -1$ for ccubic spline. 

Using the polynomial expansion 
\begin{equation}
B_{i\kappa}(z) = \sum_{j = 1}^{4}{\sum_{n = 0}^{\kappa - 1}D_{n}^{i,j}z^{n}} \text{ for } z \in (t_{i + j - 1},t_{i + j}),
\end{equation}
Where $D_{n}^{i,j}$ are the expansion coefficient for B-spline in real space.

\begin{figure}[h]
    \centering
    \includegraphics[width=0.99\columnwidth]{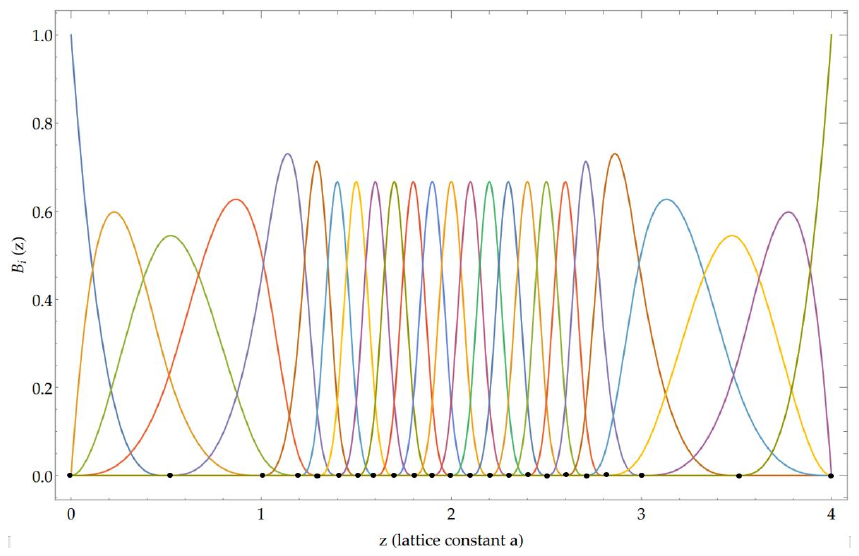}
    \caption{Cubic B-spline consisting of a 29-point knot sequence distributed over a range of $4a_{0}$ \AA{} of order $\kappa = 4$. Here, $\bullet$ denotes the knot point in the sequence.}
    \label{fig:Bspline}
\end{figure}

In reciprocal space, the Fourier transform of $B_{i\kappa}(z)$ as
\begin{equation}
\begin{split}
\widetilde{B}_{i\kappa}(g) &= \frac{1}{\sqrt{L_{c}}}\sum_{j = 1}^{4}{\sum_{n = 0}^{k}D_{n}^{i,j}}\int_{t_{i + j - 1}}^{t_{i + j}}{dz\ e^{ig(z - \frac{L_{c}}{2})}z^{n}} \\
&\equiv \frac{e^{ig(t_{i} - L_{c}/2)}}{\sqrt{L_{c}}}\sum_{j = 1}^{4}{\sum_{n = 0}^{\kappa}D_{n}^{i,j}}I_{n}^{i,j}(g),
\end{split}
\end{equation}
where $I_{n}^{i,j}(g)$ can be obtained by the following recursion relation

\begin{equation}
\begin{split}
I_{n}^{i,j}(g) &= \left. \left( \frac{z^{n}}{ig}e^{igz} \right) \right|_{0}^{\tau_{i + j} - \tau_{i}} \\
& - \frac{n}{ig}I_{j}^{n - 1}(g),
\end{split}
\end{equation}
with
\begin{equation}
I_{0}^{i,j}(g) = \left. \left( \frac{1}{ig}e^{igz} \right) \right|_{0}^{\tau_{i + j} - \tau_{i}}.
\end{equation}
Here, $L_{c}$ is the period length along $z$, and $I_{n}(g)$ can be obtained by the recursion relation.
The cubic B-spline functions used for the current calculations are shown in Fig.~\ref{fig:Bspline} 

\subsection{CONSTRUCTION OF SEPM METHOD FROM USPP CALCULATIONS}

In density functional theory (DFT), the electronic structure of a solid is addressed by solving single-particle effective Schrödinger equations, which are determined self-consistently,
\begin{equation}\label{eq:SWEdft}
\left\{ -\frac{\nabla^{2}}{2} + V_{\text{L}}(\mathbf{r}) + \hat{V}_{\text{NL}}(\mathbf{r}) + \hat{V}_{\text{SOC}} \right\} \psi_{i}(\mathbf{r}) = \varepsilon_{i} \psi_{i}(\mathbf{r})
\end{equation}
Here, \(V_{\text{L}}(r)\) denotes the effective local pseudopotential that accounts for all interactions of a single electron with its surrounding environment.
\begin{equation}
V_{\text{L}}(\mathbf{r}) = V_{\text{loc}}(\mathbf{r}) + V_{\text{HXC}}[\rho(\mathbf{r})]
\end{equation}
\begin{equation}
V_{\text{HXC}}[\rho(\mathbf{r})] = \int \frac{\rho(\mathbf{r}')}{|\mathbf{r} - \mathbf{r}'|} \, d^3\mathbf{r}' + V_{\text{XC}}[\rho(\mathbf{r})].
\end{equation}

where \(\rho(\mathbf{r}) = \sum_{i}|\psi_i(\mathbf{r})|^2\) denotes the charge density of all occupied single-particle states \(\psi_i\). First term is the Hartree (Coulomb) potential. \({V}_{\text{loc}}(\mathbf{r})\) represents the local part of the pseudopotential, \(\hat{V}_{\text{NL}}(\mathbf{r})\) represents the nonlocal part of the pseudopotential, \(\hat{V}_{\text{SOC}}(\mathbf{r})\) denotes the spin-orbit coupling term, and \(-\frac{\nabla^{2}}{2}\) represents the kinetic energy contribution.

Here,\(V_{\text{HXC}}[\rho(\mathbf{r})]\)  is not a linear combination of atomic quantities and must be determined individually for each physical system, lacking transferability between different systems. 

We will detail the procedure for constructing the semiempirical pseudopotential, which consists of three steps:

\subsubsection*{\textbf{1. Effective Local Potential Contribution}}
First, we use Density Functional Theory (DFT) \cite{ren2015mixed} to obtain the effective local potential \( V_{\text{L}}(\mathbf{r}) \). During the self-consistent cycle of the Kohn-Sham equations, the electron density is iteratively updated until the ground-state density is achieved. we perform a Fourier transform as implemented with the FFTW library \cite{frigo2005design} to convert the effective local potential from real space to reciprocal space.

\begin{equation}\label{eq:veffdft}
\begin{aligned}
    V_{\text{L}}(\mathbf{r}) &= \sum_{\sigma}\sum_{\mathbf{R}_{\sigma}} V_{a}^{\sigma}\left( \mathbf{r} - \mathbf{\tau}^{\sigma} - \mathbf{R}_{\sigma} \right) + \Delta\widetilde{V}_{\text{loc}}\left( z,G = \mathbf{0} \right) \\
&\quad + \sum_{\sigma,\mathbf{G} \neq 0}^{} \Delta\widetilde{V}_{\text{loc}}\left( z,\mathbf{G} \right)e^{i\mathbf{G}\cdot(\mathbf{\rho} - \mathbf{\tau}_{\sigma})}
\end{aligned}
\end{equation}
We will describe the detailed procedure for implementing these components  as follows:

In Eq. \ref{eq:veffdft}, the first term \(\sum_{\sigma}\sum_{\mathbf{R}_{\sigma}} V_{a}^{\sigma}\left( \mathbf{r} - \mathbf{\tau}^{\sigma} - \mathbf{R}_{\sigma} \right)\) represents the core-charge contribution to the local pseudopotential. In practical calculations, we separate the local potential of atomic PP into a long-range potential,  and a short-range remainder,

\begin{equation}\label{eq:vlocdft}
V_{a}^{\sigma}\left( \mathbf{r} \right) = V_{c,LR}^{\sigma}\left( \mathbf{r} \right) + V_{c,SR}^{\sigma}\left( \mathbf{r} \right) 
\end{equation}
can be decomposed as follows:

The first term of Eq. (\ref{eq:vlocdft}) is  corresponds to the potential due to an auxiliary charge
distribution which has been calculated analytically,
\begin{align}
V_{c,LR}^{\sigma}\left( \mathbf{r} \right) &= \int_{}^{}\frac{\rho_{c}^{\sigma}\left( \mathbf{r}^{\mathbf{'}} \right)}{\left| \mathbf{r -}\mathbf{r}^{\mathbf{'}} \right|}d\mathbf{r}^{\mathbf{'}} \\
&= \sum_{\mathbf{q},\mathbf{q}^{\mathbf{'}}}^{}{\int_{}^{}{e^{i\mathbf{q \cdot}\mathbf{(r - r}^{\mathbf{'}}\mathbf{)}}e}^{i\mathbf{q}^{\mathbf{'}}\cdot\mathbf{r}^{\mathbf{'}}}\frac{8\pi}{q^{2}}}{\widetilde{\rho}}_{c}^{\sigma}\left( \mathbf{q}^{\mathbf{'}} \right)d\mathbf{r}^{\mathbf{'}} \nonumber \\
&= \sum_{\mathbf{q}}^{}{\int_{}^{}e^{i\mathbf{q \cdot}\mathbf{r}}\frac{8\pi}{q^{2}}{\widetilde{\rho}}_{c}(\mathbf{q})} \nonumber \\
&= 8\pi\sum_{\mathbf{q}_{\parallel},q_{z}}^{}{\int_{}^{}e^{i\mathbf{q \cdot}\mathbf{r}}\frac{1}{q_{t}^{2} + q_{z}^{2}}{\widetilde{\rho}}_{c}^{\sigma}(\mathbf{q}_{\parallel}+q_{z}\widehat{\mathbf{z}})} \nonumber
\end{align}
where \(\widetilde{\rho}_{c}\left( \mathbf{q} \right) = \frac{Z_{c}}{\mathrm{\Omega}}e^{- q^{2}R_{c}^{2}/4}\) (\(\mathrm{\Omega} = \) sample volume).

The Fourier transform of \(V_{c,LR}^{\sigma}\left( \mathbf{r} \right)\) in the 2D plane is given by
\begin{equation}\label{eq:FTlocLR}
V_{c,LR}^{\sigma}\left( z,\mathbf{g}_{\parallel} \right) = \frac{8\pi}{\mathrm{\Omega}_{c}}\sum_{g_{z}}^{}e^{ig_{z} \cdot z}\frac{Z_{c}}{\mathbf{g}_{\parallel}^{2} + g_{z}^{2}}e^{- {\left( \mathbf{g}_{\parallel}^{2} + g_{z}^{2} \right)R}_{c}^{2}/4}
\end{equation}
(\(\mathrm{\Omega}_{c} = A_{c}L_{c} =\) supercell volume)

\begin{equation}
\begin{aligned}
&\left\langle \mathbf{K};B_{i} \middle| V_{c,LR}^{\sigma} \middle| \mathbf{K'};B_{i'} \right\rangle = \int_{}^{}{dzB_{i}(z)}{\widetilde{V}}_{c,LR}^{\sigma}\left( z,\mathbf{G}^{\prime}-\mathbf{G} \right)B_{i'}(z) \\
&= \frac{8\pi}{\mathrm{\Omega}_{c}}\sum_{g_{z}}^{}{I_{ii'}(g_{z})}\frac{Z_{c}}{\mathbf{\Delta G}^{2} + g_{z}^{2}}e^{- {\left( \mathbf{\Delta G}^{2} + g_{z}^{2} \right)R}_{c}^{2}/4}
\end{aligned}
\end{equation}

where
\(I_{ii'}(g_{z}) = \left\langle B_{i} \middle| e^{ig_{z} z} \middle| B_{i'} \right\rangle\) and
\(\mathbf{\Delta G} = \mathbf{G}^{\prime} - \mathbf{G}\).


For the short-range Core contribution part \(V_{c,SR}^{\sigma}\left( \mathbf{r} \right)\)  of Eq. \ref{eq:vlocdft}  can be well fitted by the following function:
\begin{widetext}
\begin{equation} \label{eq:SRcore}
V_{c,SR}^{\sigma}\left( \mathbf{r} \right) = \begin{cases}
(C_{0,1}^{M} + Cr^{2} + C_{2,1}^{M}r^{4} + C_{3,1}^{M}r^{6} + B_{4,1}^{M}r^{8})\Theta\left( r_{1}^{M} - r \right) \\
\quad + \left( C_{0,2}^{M} + C_{1,2}^{M}r^{2} + C_{2,2}^{M}r^{4} + C_{3,2}^{M}r^{6} + C_{4,2}^{M}r^{8} \right)e^{- \alpha_{M}r^{2}}, & \text{M atom} \\[10pt]
\sum_{i = 1}^{3}\left( C_{0,i}^{X} + C_{1,i}^{X}r^{2} + C_{2,i}^{X}r^{4} + C_{3,i}^{X}r^{6} + C_{4,i}^{X}r^{8} + C_{5,i}^{X}r^{10} \right)\Theta\left( r_{i}^{X} - r \right)\Theta\left( r - r_{i - 1}^{X} \right), & \text{X atom}
\end{cases}
\end{equation}
\end{widetext}
where \(r_{1}^{M} \) denotes the cutoff radius for the first term in the M-atom pseudopotential;
\({r_{0}^{X},r}_{1}^{X},r_{2}^{X}\), and
\(r_{3}^{X}\) denote the cutoff radii for the three terms in the X-atom pseudopotential.
Similarly, 
\begin{equation}
V_{c,SR}^{\sigma}\left( z,\mathbf{g}_{\parallel} \right) = \sum_{g_{z}}^{}e^{ig_{z} \cdot z}\widetilde{V}_{va}(\mathbf{g}_{\parallel},\mathbf{g}_{z})
\end{equation}
With
\begin{equation}
\widetilde{V}_{c,SR}^{\sigma}\left( \mathbf{g}_{\parallel},\mathbf{g}_{z} \right) = \int_{}^{}{rdr\ \sin\left( |\mathbf{g}_{\parallel}+\mathbf{g}_{z}|r \right)}\frac{{V}_{c,SR}^{\sigma}}{|\mathbf{g}_{\parallel}+\mathbf{g}_{z}|}.
\end{equation}
We have used the expansion
\begin{equation}
e^{-iK\cdot \mathbf{r}} = 4\pi\sum_{lm}^{}{(-i)^{l}j_{l}(Kr)Y_{lm}\left( \Omega_{K} \right)Y_{lm}^{\ast}(\Omega)}
\end{equation}

and \(j_{0}(Kr) = \frac{\sin(Kr)}{Kr}\) with \(\mathbf{K} = \mathbf{g}_{\parallel} + \mathbf{g}_{z}\). The matrix elements of \(V_{c,SR}^{\sigma}\left( \mathbf{r} \right)\) in the 3D reciprocal basis \(\left| \mathbf{K} \right\rangle = \left| \mathbf{k} + \mathbf{G} + \mathbf{g}_{z} \right\rangle\) are

\begin{equation}
\begin{split}
&\sum_{\sigma,\mathbf{R}}^{}{}\left\langle \mathbf{k} + \mathbf{G} + \mathbf{g}_{z} \middle| V_{a}^{\sigma}\left( \mathbf{r} - \boldsymbol{\tau}^{\sigma} - \mathbf{R} \right) \middle| \mathbf{k} + \mathbf{G}' + \mathbf{g}'_{z} \right\rangle \\
&= \widetilde{V}_{va}\left( \boldsymbol{\Delta} \mathbf{G} + \boldsymbol{\Delta} \mathbf{g}_{z} \right) \\
&= e^{i\left( \boldsymbol{\Delta} \mathbf{G} + \boldsymbol{\Delta} \mathbf{g}_{z} \right) \cdot \boldsymbol{\tau}_{\sigma}}\int_{}^{}{rdr\ \sin\left( \left| \boldsymbol{\Delta} \mathbf{G} + \boldsymbol{\Delta} \mathbf{g}_{z} \right|r \right)}\frac{{V}_{c,SR}^{\sigma}}{\left| \boldsymbol{\Delta} \mathbf{G} + \boldsymbol{\Delta} \mathbf{g}_{z} \right|}.
\end{split}
\end{equation}
where we have used the expansion
\(z_{j} = (z \pm \tau_{z})\) for the X atoms with \(j = 1,2\). Note that we have carried out the 2D Fourier transform for the M atom analytically. For the X atoms, we can evaluate the \(\rho\) dependent part on an FFT mesh within the 2D unit cell analytically and then use the 2D FFT to obtain \({V}_{c,SR}^{\sigma}\left( z,\mathbf{G} \right)\). \(\{\mathbf{R}_{\mu}; \mu = 1,\ldots,7\}\) denote the centers of five unit cells included in the evaluation of \({V}_{c,SR}^{\sigma}\left( z,\mathbf{G} \right)\). Since the fitting function is truncated for \(\left| \mathbf{r} - \mathbf{R}_{\mu} \right| > r_{c}\), the X-atom contribution to \({V}_{c,SR}^{\sigma}\left( z,\mathbf{G} \right)\) only comes from the central cell and six neighbors.

The term \(V_{\text{HXC}}[\rho(\mathbf{r})]\)is not a linear combination of atomic quantities and must be determined independently for each physical system, as it lacks transferability between different systems. The last two terms in Eq. \ref{eq:veffdft} correspond to the isotropic \([G = 0]\) and anisotropic \([G \neq 0]\) components of \(V_{\text{HXC}}[\rho(\mathbf{r})]\). \(\Delta\widetilde{V}_{\text{loc}}\left( z,G = \mathbf{0} \right)\) denotes the Fourier transform of the correction to the total
local potential \(V_{loc}\left( \mathbf{r} \right)\) in the 2D reciprocal space which includes the Hartree potential due to the valence
charges and the exchange-correlation potential. The isotropic main term (with \(\mathbf{G = 0}\) ) is parameterized in terms of three spherical
Gaussian functions centered at each atomic site. We have
\begin{equation}\label{eq:VHXC_G0}
\begin{aligned}
\Delta{\widetilde{V}}_{\text{loc}}\left( z,\mathbf{0} \right) = 
&\sum_{j = 1,2} \left[ A_{j}^{M}e^{- \alpha_{j}^{M}\left( z - Z^{M} \right)^{2}} + \right. \\
&\left. A_{j}^{X}\left( e^{- \alpha_{j}^{X}\left( z - Z_{1}^{X} \right)^{2}} + e^{- \alpha_{j}^{X}\left( z - Z_{2}^{X} \right)^{2}} \right) \right]
\end{aligned}
\end{equation}
where \(A_{j}^{w}\), \(A_{j}^{\text{se}}\), \(\alpha_{j}^{w}\), and
\(\alpha_{j}^{\text{se}}\) are determined by fitting the spatial average of \(V_{\text{loc}}\left( \mathbf{r} \right)\) in the 2D plane as
a function of \(z\). \(Z^{M}\) denotes the \(z\) position of the M atom.
\(Z_{j}^{X}\) \((j = 1,2)\) denotes the \(z\) positions of the two X atoms.

The last term 
\(
\sum_{\sigma,\mathbf{G} \neq 0}^{} \Delta\widetilde{V}_{\text{loc}}\left( z,\mathbf{G}\right)e^{i\mathbf{G}\cdot(\mathbf{\rho} - \mathbf{\tau}_{\sigma})}
\)
in Eq. \ref{eq:veffdft} represents the difference 
\begin{align} \label{eq:VHXC}
\Delta \widetilde{V}_{\text{HXC}}\left( z,\mathbf{G \neq 0} \right) = &
V_{\text{L}}(\mathbf{r}) - \sum_{\sigma} \sum_{\mathbf{R}_{\sigma}} V_{a}^{\sigma}\left( \mathbf{r} - \mathbf{\tau}^{\sigma} - \mathbf{R}_{\sigma} \right) \nonumber \\
& - \Delta \widetilde{V}_{\text{loc}}\left( z, \mathbf{G} = \mathbf{0} \right)
\end{align}
which is anisotropic in nature (\(\mathbf{G} \neq \mathbf{0}\)). The isotropic part is already well fitted by \(\Delta{\widetilde{V}}_{\text{loc}}\left( z,\mathbf{0} \right)\). The reciprocal lattice vector \(\mathbf{G}\) can be sorted into many shells, with the magnitude of \(\mathbf{G}\) vectors being the same for each shell.
The contributions of M and X atoms to the structure factor of the crystal structure \(\sigma(MX_{2})\), are given by:
\begin{equation} \label{eq:strucfact}
\left\{
\begin{aligned}
    S^{M}\left( \mathbf{G} \right) &= e^{-i\mathbf{G} \cdot \mathbf{\tau}^{M}} \\
    S^{X}\left( \mathbf{G} \right) &= e^{-i\mathbf{G} \cdot \mathbf{\tau}^{X}}
\end{aligned}
\right.
\end{equation}
We note that both X atoms share the same in-plane coordinate. To fit the differences arising in Eq. \ref{eq:VHXC}, we divide the contributions of \(V_{\text{HXC}}[\rho(\mathbf{r})]\) into long-range (LR) and short-range (SR) components. We express this as:
\begin{equation}\label{eq:difVHXC}
  \Delta{\widetilde{V}}_{\text{HXC}}\left( z, \mathbf{G} \neq 0 \right) = \Delta{\widetilde{V}}_{\text{HXC,LR}}\left( z, \mathbf{G} \right) + \Delta{\widetilde{V}}_{\text{HXC,SR}}\left( z, \mathbf{G} \right)  
\end{equation}

The long-range (LR) contribution is expected to become negligible for higher \(\mathbf{G}\) shells (\textgreater 3.1 a.u.). Therefore, we first determine the short-range (SR) component by fitting the DFT result (Eq. \ref{eq:veffdft}) for shells with \(\mathbf{G}\) values in the higher range (\textgreater 3.1 a.u.). The \(z\)-dependence of \(\Delta{\widetilde{V}}_{\text{SR}}\left( z, \mathbf{G} \right)\) can be well-fitted by a single Gaussian function (centered at \(z = 0\)) of the form:
\begin{equation}\label{eq:difVHXCSR}
\Delta{\widetilde{V}}_{\text{HXC,SR}}\left( z, \mathbf{G} \right) = D G^{6} e^{- b G^{2}} e^{- c z^{2}}
\end{equation}

It is observed that the X-atoms contribution is negligible for higher \(\mathbf{G}\) shells.

Then the remaining long-range function for \(\mathbf{G}\) shells (\(\leq 3.1\) a.u.), \(\Delta{\widetilde{V}}_{\text{HXC,LR}}\left( z, \mathbf{G} \right)\) in Eq. \ref{eq:difVHXC}, is by fitting the points for given structure \(\sigma\) of the form:
\begin{equation}
\begin{aligned}
\Delta{\widetilde{V}}_{\text{HXC,LR}}\left( z,\mathbf{G} \right) = & f_{LR}^{M}(z) S^{M}\left( \mathbf{G} \right) \\
& + \left( f_{LR}^{X}(z) + f_{LR}^{BC}(z) \right) S^{X}\left( \mathbf{G} \right)
\end{aligned}
\end{equation}

where
\begin{equation}
\begin{split}
f_{LR}^{M}(z) &= \sum_{i = 1,2,3} A_{i}^{M} e^{- \alpha_{i}^{M} \left( z - \tau^{M} \right)^{2}} \\
f_{LR}^{X}(z) &= \sum_{i = 1,2,3} A_{i}^{X} \left[ e^{- \alpha_{i}^{X} \left( z - \tau^{X} \right)^{2}} + e^{- \alpha_{i}^{X} \left( z + \tau^{X} \right)^{2}} \right] \\
f_{LR}^{BC}(z) &= \sum_{i = 1,2,3} A_{i}^{BC} e^{- \alpha_{i}^{BC} \left( z - \tau^{M} \right)^{2}}
\end{split}
\end{equation}
Here, \( f_{LR}^{M} \), \( f_{LR}^{X} \), and \( f_{LR}^{BC} \) represent the shape functions for the long-range part of \(\mathbf{G}\) shells corresponding to the M atom, X atom, and bond-charge, respectively. \( A_{i}^{M}, A_{i}^{X}, \alpha_{i}^{M}, \) and \( \alpha_{i}^{X} \) are the fitting parameters for each shell in the crystal structure \(\sigma\).

Finally, the total effective local potential due to the contribution of the ionic, Hartree and XC parts after correction in our SEPM  is given by
\begin{equation}
\begin{aligned}
V_{\text{loc}}\left( z, \mathbf{G} \right) &= V_{c,LR}^{\sigma}\left( z, \mathbf{g}_{\parallel} \right) + V_{c,SR}^{\sigma}\left( z, \mathbf{G} \right) + \Delta{\widetilde{V}}_{\text{loc}}\left( z, \mathbf{0} \right)  + \\
& \quad \sum_{s = 1,\ldots,5} \Delta{\widetilde{V}}_{HXC,LR}^{s} \left( z, \mathbf{G} \right) 
 +\Delta{\widetilde{V}}_{\text{HXC,SR}}\left( z, \mathbf{G} \right)
\end{aligned}
\end{equation}

\subsubsection*{\textbf{2. Nonlocal Pseudopotential contribution}}

Nonlocal pseudopotentials (NLPPs) are crucial for accurately describing the electronic structures TMDCs, which include  transition metals. These materials present challenges due to the significant influence of core electrons on their electronic properties.  Unlike spherically symmetric local pseudopotentials, NLPPs vary with angular momentum. Incorporating both local and NLPPs can provide a more precise description of the energy range for valence band edges and better alignment with density functional theory (DFT) calculations or experimental data. In the scalar-relativistic (SR) approximation, the separable formulation of
Kleinman and Bylanderis \cite{kleinman1982efficacious} (KB) given by
\begin{equation}\label{eq:nonlocalpot}
\widehat{V}_{nl} = \sum_{\sigma lm, nn'} E_{lm}^{nn'} \left| \beta_{lm}^{n\sigma} \right\rangle \left\langle \beta_{lm}^{n'\sigma} \right|,
\end{equation}
Here, projector functions \(\beta_{lm}^{n\sigma}\) as well as coefficient \(E_{lm}^{nn'}\) characterise the pseudopotential and their values different for different atomic species (Mo, W, S, Se). The
best-fit \(\beta\) functions take the following form:
\begin{equation}
\beta_{lm}^{n\sigma} \left( \mathbf{r} \right) = A_{l}^{n\sigma}(r) r^{l} Y_{lm}(\Omega).
\end{equation}
Here, \(A_{l}^{n\sigma}(r)\) is the energy dependent well depth. We fit the beta function for each  orbital \(|l,m\rangle\) for different atoms using  ultrasoft pseudopotential  \cite{vanderbilt1990soft,garrity2014pseudopotentials} implemented in DFT code \cite{ren2015mixed}. 
For efficient fitting, we break the beta functions into two segments with a threshold \( R_{s} \) between the starting point of the first segment and \( R_{cut} \).

For the first segment \( r < R_{s} \), the fitting parameter used in our SEPM  for \( A_{l}^{n\sigma}(r) \) is a polynomial times a Gaussian function:
\begin{equation}\label{eq:nonlocalseg1}
   A_{l}^{n\sigma}(r) = \sum_{p=0}^{5} C_{p} \mathbf{r}^{2p} e^{-\alpha_{p}\mathbf{r}^{2}}  
\end{equation}

For the second segment \( r \geq R_{s} \), the fitting function is of the form:
\begin{equation}\label{eq:nonlocalseg2}
   A_{l}^{n\sigma}(r) = \sum_{p=0}^{5} C_{p} \mathbf{r}^{2p} e^{-\alpha_{p}\mathbf{r}^{2}} + C_{6} \mathbf{r} e^{-\alpha_{6}\mathbf{r}^{2}}
\end{equation}
where \( C_{p} \) and \( \alpha_{p} \) are the coefficients of the polynomials for different atoms \( \sigma \), and \( \alpha_{p} \) is the coefficient of the exponential term.

The matrix elements for NLPPs are given by[]
\begin{equation}\label{eq:matNLPP}
\begin{aligned}
& \left\langle \mathbf{K};B_{i} \middle| \widehat{V}_{nl} \middle| \mathbf{K}';B_{i'} \right\rangle \\
& = \sum_{\sigma lm,nn'} E_{lm}^{nn'} 
\left\langle \mathbf{K};B_{i} \middle| \beta_{lm}^{n\sigma} \right\rangle 
\left\langle \beta_{lm}^{n'\sigma} \middle| \mathbf{K}';B_{i'} \right\rangle 
e^{i(\mathbf{G}' - \mathbf{G}) \cdot \mathbf{\tau}_{\sigma}} \\
& = \frac{1}{A_{c}} \sum_{\sigma l,m \geq 0} \sum_{n,n'} E_{lm}^{nn'} 
\operatorname{Re}\left[ P_{lm}^{in\sigma}(\mathbf{K}) 
P_{lm}^{\sigma i'n'\sigma \ast}(\mathbf{K}') \right] 
e^{i(\mathbf{G}' - \mathbf{G}) \cdot \mathbf{\tau}_{\sigma}}
\end{aligned}
\end{equation}
where, $\mathbf{K} = \mathbf{k} + \mathbf{G}$ and $\mathbf{K}' = \mathbf{k} + \mathbf{G}'$.
In terms of 3D plane waves, indexed by \(\mathbf{Q} = \mathbf{K} + g_{z} \widehat{\mathbf{z}}\) (where \(g_{z}\) denotes the reciprocal lattice vectors along the z-axis for the supercell adopted), the projection of the beta function in the mixed basis (with respect to the position of an atom \(\sigma\)) is given by
\begin{equation}\label{eq:matPlm}
\begin{split}
P_{lm}^{in\sigma}\left( \mathbf{K} \right) = & \, i^{l} \sqrt{A_{c}} \left\langle \mathbf{K}; B_{i} \middle| \beta_{lm}^{n\sigma} \right\rangle \\
= & \, \frac{1}{\sqrt{L_{c}}} \sum_{\mathbf{g}_{z}} \widetilde{B}_{i}\left( g_{z} \right) I_{l}(Q) Y_{lm}(\widehat{\mathbf{Q}}).
\end{split}
\end{equation}

Here, \(\widetilde{B}_{i}(g_{z})\) is the Fourier transform of \(B_{i}(z)\) as given in eq(?).
\begin{equation}
\begin{split}
I_{l}(Q) &= 4\pi\int_{0}^{R_{C}} drr^{l + 2}A_{l}(r)j_{l}(Qr) \\
&= 4\pi\sum_{i} A_{l}(r_{i})\left( r^{l + 2}j_{l + 1}(Qr) \right)\bigg|_{r_{i -}}^{r_{i +}}
\end{split}
\end{equation}
as
\begin{equation}
\begin{split}
   \int_{r_{i -}}^{r_{i +}} drr^{l + 2}j_{l}(Qr) &= \sqrt{\frac{\pi}{2}}\int_{r_{i -}}^{r_{i +}} drr^{l + 2 - 1/2}J_{l + 1/2}(Qr) \\
  & = \left( r^{l + 2}j_{l + 1}(Qr) \right)|_{r_{i -}}^{r_{i +}}/Q.   
\end{split}
\end{equation}
$r_{i}$ denotes the midpoint and $r_{i -}$ $(r_{i +})$ denotes the lower (upper) bound of the $i$-th interval.
 We can rewrite Eq. (\ref{eq:matNLPP}) as
\begin{equation}
\begin{split}
    & \left\langle \mathbf{K};B_{i} \middle| {\widehat{V}}_{nl} \middle| \mathbf{K}';B_{i'} \right\rangle \\
    &= \frac{1}{A_{c}}\sum_{\sigma l,m \geq 0}^{}{\sum_{n,n'}^{\ }E_{lm}^{nn'} \operatorname{Re}\left\lbrack P_{lm}^{in\sigma}\left( \mathbf{K} \right)P_{lm}^{{\sigma i}^{'}n^{'}\sigma \ast}\left( \mathbf{K}^{'} \right) \right\rbrack{d_{m}e}^{i(\mathbf{G}^{'} - \mathbf{G}) \cdot \mathbf{\tau}_{\sigma}}}
\end{split}
\end{equation}

where \(d_{0} = 1\) and \(d_{m} = 2\) for \(|m| > 0\). We define
\(Q_{lm}^{in\sigma}(K) = \left| P_{lm}^{in\sigma}\left( \mathbf{K} \right) \right|\)
and get

\(P_{lm}^{in\sigma}\left( \mathbf{K} \right) = Q_{lm}^{in\sigma}(K)e^{im\varphi_{K}} = Q_{lm}^{in\sigma}(K)\left( \frac{K_{x} + iK_{y}}{K} \right)^{m}.\)
Therefore, we can write
\begin{equation}
\begin{split}
     \left\langle \mathbf{K};B_{i} \middle| \widehat{S} \middle| \mathbf{K}';B_{i'} \right\rangle &= \sum_{n,n',l,|m|}^{}{q_{l}^{nn'}Q_{lm}^{in}(K)Q_{lm}^{i'n'}(K')} \\
     &\quad \times \left( \frac{{K_{x}K'}_{x} + K_{y}{K'}_{y}}{KK'} \right)^{|m|} d_{m} S^{\sigma}(\Delta\mathbf{G})  
\end{split}
\end{equation}
where \(S^{\sigma}\)(\(\Delta\mathbf{G})\) is the \(\sigma\)-atom
contribution to the structure factor given in Eq. (\ref{eq:strucfact}). Similarly, for
the nonlocal pseudopotential, we have
\begin{align}
\begin{split}
\left\langle \mathbf{K};B_{i} \middle| \widehat{V}_{nl} \middle| \mathbf{K}';B_{i'} \right\rangle = & \sum_{\sigma,n,n',l,|m|}^{}{E_{l}^{nn'}Q_{lm}^{in}(K)Q_{lm}^{i'n'}(K')} \\
& \times \left( \frac{{K_{x}K'}_{x} + K_{y}{K'}_{y}}{KK'} \right)^{|m|}d_{m}S^{\sigma}(\Delta\mathbf{G})
\end{split}
\end{align}
By obtaing the magnitude of \(E_{l}^{nn'}\) and \(q_{l}^{nn'}\), we can obtain the matrix element of NLPPs for \( \widehat{V}_{nl}\) and \(\widehat{S}\).
\subsubsection*{\textbf{3. Spin-Orbit Coupling contribution}}
The incorporation of relativistic effects, which arise from the coupling between the orbital angular momentum of electrons and their spin can be integrated into the formulation of ultrasoft pseudopotentials.
For fully relativistic (FR) pseudopotential \cite{corso2005spin}, we replace the
\(\beta_{lm}^{n\sigma}\left( \mathbf{r} \right)\) function by the following spinors. Here's the equation with the split environment:
\begin{equation}
\begin{aligned}
    & \beta_{l + 1/2,m}^{n\sigma}(\mathbf{r}) = \\
    &A_{l + 1/2}^{n\sigma}(r)r^{l}\left( c_{lm}^{+}Y_{l,m}(\Omega)\chi_{1/2} + c_{lm}^{-}Y_{l,m + 1}(\Omega)\chi_{- 1/2} \right) \\
    & \beta_{l - 1/2,m}^{n\sigma}(\mathbf{r}) = \\
    &A_{l - 1/2}^{n\sigma}(r)r^{l}\left( d_{lm}^{+}Y_{l,m - 1}(\Omega)\chi_{1/2} + d_{lm}^{-}Y_{l,m}(\Omega)\chi_{- 1/2} \right) 
\end{aligned}
\end{equation}
where \(c_{lm}^{+} = \sqrt{\frac{l + m + 1}{2l + 1}}\),
\(c_{lm}^{-} = \sqrt{\frac{l - m}{2l + 1}}\),
\(\ d_{lm}^{+} = \sqrt{\frac{l - m + 1}{2l + 1}}\),
\(d_{lm}^{-} = - \sqrt{\frac{l + m}{2l + 1}}\) and \(\chi_{\pm 1/2}\)
denoting the up and down spin components of the basis states.

The projection function \(P_{lm}^{in\sigma}\left( \mathbf{K} \right)\) in Eq. (\ref{eq:matPlm}) is replaced by
\begin{equation}
\left\{
\begin{aligned}
&\left\langle \mathbf{K};B_{i};\chi_{\pm 1/2} \middle| \beta_{l + 1/2,m}^{n\sigma} \right\rangle = P_{l + 1/2,m}^{in\sigma \pm}\left( \mathbf{K} \right)e^{- i\mathbf{G} \cdot \mathbf{\tau}^{\sigma}} \\
&\left\langle \mathbf{K};B_{i};\chi_{\pm 1/2} \middle| \beta_{l - 1/2,m}^{n\sigma} \right\rangle = P_{l - 1/2,m}^{in\sigma \pm}\left( \mathbf{K} \right)e^{- i\mathbf{G} \cdot \mathbf{\tau}^{\sigma}}
\end{aligned}
\right.
\end{equation}

\begin{equation}
\left\{
\begin{array}{@{}r@{}}
\begin{aligned}
&P_{l + 1/2,m}^{in\sigma +}\left( \mathbf{K} \right) = c_{lm}^{+}\frac{1}{\sqrt{L_{c}}}\sum_{\mathbf{g}_{z}}^{}{\widetilde{B}_{i}\left( g_{z} \right)}I_{l + 1/2}(Q)Y_{lm}(\widehat{\mathbf{Q}}) \\
&= c_{lm}^{+}P_{l + ,m}^{in\sigma}\left( \mathbf{K} \right) \\
\\
&P_{l + 1/2,m}^{in\sigma -}\left( \mathbf{K} \right) = c_{lm}^{-}\frac{1}{\sqrt{L_{c}}}\sum_{\mathbf{g}_{z}}^{}{\widetilde{B}_{i}\left( g_{z} \right)}I_{l + 1/2}(Q)Y_{lm + 1}(\widehat{\mathbf{Q}}) \\
&= c_{lm}^{-}P_{l + ,m + 1}^{in\sigma}\left( \mathbf{K} \right)\quad \text{(for } m < l\text{)}
\end{aligned} \\
\\
\begin{aligned}
&P_{l - 1/2,m}^{in\sigma +}\left( \mathbf{K} \right) = d_{lm}^{+}\frac{1}{\sqrt{L_{c}}}\sum_{\mathbf{g}_{z}}^{}{\widetilde{B}_{i}\left( g_{z} \right)}I_{l - 1/2}(Q)Y_{lm - 1}(\widehat{\mathbf{Q}}) \\
&= d_{lm}^{+}P_{l - ,m - 1}^{in\sigma}\left( \mathbf{K} \right) \quad \text{(for } m > - l\text{)}
\end{aligned} \\
\\
\begin{aligned}
&P_{l - 1/2,m}^{in\sigma -}\left( \mathbf{K} \right) = d_{lm}^{-}\frac{1}{\sqrt{L_{c}}}\sum_{\mathbf{g}_{z}}^{}{\widetilde{B}_{i}\left( g_{z} \right)}I_{l - 1/2}(Q)Y_{lm}(\widehat{\mathbf{Q}}) \\
&= d_{lm}^{-}P_{l - ,m}^{in\sigma}\left( \mathbf{K} \right)
\end{aligned}
\end{array}
\right.
\end{equation}
with 
\begin{equation}
  \begin{split}
I_{l \pm 1/2}(Q) &= 4\pi\int_{0}^{R_{C}} drr^{l + 2}A_{l \pm 1/2}(r)j_{l}(Qr) \\
&= 4\pi\sum_{i} A_{l \pm 1/2}(r_{i})\left( r^{l + 2}j_{l + 1}(Qr) \right)|_{r_{i +}}^{r_{i +}}
\end{split}  
\end{equation}
Note: \(P_{l \pm ,m}^{in\sigma}\left( \mathbf{K} \right)\) is obtained
form \(P_{l,m}^{in\sigma}\left( \mathbf{K} \right)\) in Eq. (\ref{eq:matPlm}) with \(I_{l}(Q)\) replaced by \(I_{l \pm 1/2}(Q)\).
Then we get FR matrix element for nonlocal pseudopotentials
\begin{equation}
\begin{split}
& \left\langle \mathbf{K};B_{i};\chi_{s} \middle| {\widehat{V}}_{nl} \middle| \mathbf{K}^{'};B_{i^{'}};\chi_{s'} \right\rangle  \\
& =\sum_{\sigma lm \pm ,nn^{'}} E_{lm}^{nn^{'}} \left\langle \mathbf{K};B_{i};\chi_{s} \middle| \beta_{l \pm 1/2m}^{n\sigma} \right\rangle \left\langle \beta_{l \pm 1/2m}^{n^{'}\sigma} \middle| \mathbf{K}^{'};B_{i^{'}};\chi_{s'} \right\rangle \\
&\quad \times e^{i(\mathbf{G}^{'} - \mathbf{G}) \cdot \mathbf{\tau}_{\sigma}} \\
&= \frac{1}{A_{c}}\sum_{\sigma lm \pm} \sum_{nn'} E_{lm}^{nn'} P_{l \pm 1/2,m}^{in\sigma s}(\mathbf{K}) \\
& \quad \times P_{l \pm 1/2,m}^{i^{'}n^{'}\sigma s' *}(\mathbf{K}') e^{i(\mathbf{G}^{'} - \mathbf{G}) \cdot \mathbf{\tau}_{\sigma}} \\
&= \frac{1}{A_{c}}\sum_{\sigma lm \pm} \sum_{nn'} E_{lm}^{nn'} Q_{l \pm 1/2,m}^{in\sigma s}(K) Q_{l \pm 1/2,m}^{i^{'}n^{'}\sigma s'}(K') \\
& \quad \times e^{im_{s \pm}\varphi_{K}}e^{- im_{s' \pm}\varphi_{K'}}e^{i(\mathbf{G}^{'} - \mathbf{G}) \cdot \mathbf{\tau}_{\sigma}}
\end{split}
\end{equation}
Here
\(Q_{l \pm 1/2,m}^{in\sigma s}(K) = \left| P_{l \pm 1/2,m}^{in\sigma s}\left( \mathbf{K} \right) \right|\)
and \(m_{s \pm} = m \pm 1/2 - s\) for \emph{s=}1/2 and -1/2. Similarly,
we have
\begin{equation}
\begin{split}
& \left\langle \mathbf{K};B_{i};\chi_{s} \middle| \widehat{S} \middle| \mathbf{K}^{'};B_{i^{'}};\chi_{s'} \right\rangle \\
&= \sum_{\sigma lm \pm ,nn^{'}} q_{lm}^{nn^{'}} \left\langle \mathbf{K};B_{i};\chi_{s} \middle| \beta_{l \pm 1/2m}^{n\sigma} \right\rangle \\
& \quad \times \left\langle \beta_{l \pm 1/2m}^{n^{'}\sigma} \middle| \mathbf{K}^{'};B_{i^{'}};\chi_{s'} \right\rangle e^{i(\mathbf{G}^{'} - \mathbf{G}) \cdot \mathbf{\tau}_{\sigma}}
\end{split}
\end{equation}
The projection of the beta function in a mixed basis, relative to the position \( \tau_{\sigma, z} \) of an atom along the \( z \)-axis in the supercell, is given by:

Since \( B_{i\kappa}(z) = \sum_{g_{z}}^{}{{\widetilde{B}}_{i\kappa}(g)}e^{ig_{z}z} \) is real, it follows that \( {\widetilde{B}}_{i}( - g) = {\widetilde{B}}_{i}^{\ast}(g) \). Consequently, we obtain:
\begin{equation}
\begin{split}
&\sum_{\mathbf{g}_{z}}^{}{{\widetilde{B}}_{i}\left( g_{z} \right)e^{ig_{z}\tau_{\sigma z}}} \\
&\quad \times \left\{ Y_{00}\left( \widehat{\mathbf{Q}} \right), Y_{11}\left( \widehat{\mathbf{Q}} \right), Y_{20}\left( \widehat{\mathbf{Q}} \right), Y_{22}\left( \widehat{\mathbf{Q}} \right) \text{ even in } g_{z} \right\} \\
&= {\widetilde{B}}_{i}(0) \left\{ Y_{00}(\widehat{\mathbf{K}}), Y_{11}(\widehat{\mathbf{K}}), Y_{20}(\widehat{\mathbf{K}}), Y_{22}\left( \widehat{\mathbf{K}} \right) \right\} \\
&\quad + 2\sum_{\mathbf{g}_{z > 0}}^{}{\operatorname{Re}\left[ {\widetilde{B}}_{i}\left( g_{z} \right)e^{ig_{z}\tau_{\sigma z}} \right]} \\
&\quad \times \left\{ Y_{00}(\widehat{\mathbf{Q}}), Y_{11}(\widehat{\mathbf{Q}}), Y_{20}(\widehat{\mathbf{Q}}), Y_{22}\left( \widehat{\mathbf{Q}} \right) \right\}
\end{split}
\end{equation}

\begin{equation}
\begin{split}
&\sum_{\mathbf{g}_{z}}^{}{{\widetilde{B}}_{i}\left( g_{z} \right)e^{ig_{z}\tau_{\sigma z}}} \{ Y_{10}\left( g_{z} \right), Y_{21}\left( g_{z} \right) \} \\
&= 2i\sum_{\mathbf{g}_{z > 0}}^{}{\operatorname{Im}\left[ {\widetilde{B}}_{i}\left( g_{z} \right)e^{ig_{z}\tau_{\sigma z}} \right]} \{ Y_{10}\left( g_{z} \right), Y_{21}\left( g_{z} \right) \} \text{ (odd in } g_{z})
\end{split}
\end{equation}
\section{\textbf{Application on Layer Transition-Metal Dichalcogenides }}
In this section, we apply the SEPM approach described in Sec. II to monolayer and bilayer MoS\textsubscript{2}, WS\textsubscript{2}, MoSe\textsubscript{2}, and WSe\textsubscript{2} system.

\subsection{\textbf{Calculating SEPM pseudopotential for Monolayer TMDCs }}

The starting point of our SEPM potential involves solving and extracting the local part of the self-consistent effective potential Eq.(\ref{eq:SWEdft}) using our well-developed mixed basis DFT code \cite{ren2015mixed}, which employs plane waves for the in-plane direction and B-splines for the out-of-plane \(z\)-direction. The generalized gradient approximation with the Perdew-Burke-Ernzerhof (PBE) exchange-correlation functional \cite{perdew1996generalized} is adopted for all crystal structures considered. We found that an energy cutoff of 20 Ry for 2D plane waves and 25 for B-splines, distributed over a range of 4.0 \(a_{0}\), is sufficient for convergence, where \(a_{0}\) is the lattice constant of the respective structure. A Monkhorst-Pack \(7 \times 7\) mesh, including the \(\Gamma\) point, is used to sample the 2D irreducible Brillouin zone (IBZ) \cite{monkhorst1976special}. Core and valence electrons are treated using the ultrasoft pseudopotential method (USPP) \cite{vanderbilt1990soft,garrity2014pseudopotentials,schlipf2015optimization}. The USPP potentials represent the nuclei plus core electrons up through the 4d shell for Mo and up through the 5d shell for W for TMDC layer materials \cite{wilson1969transition}. For chalcogens, the s and p electrons of the outermost shell are treated as valence. For transition metal Mo and W, we employ non-local core correction \cite{louie1982nonlinear}. The potential is determined self-consistently until its change is less than \(10^{-7}\) Ry.

First, we extract long range part of core contribution of potential and solve analytically as first term of Eq. (\ref{eq:vlocdft}). For remaining short range part of local potential, we fitted for M (Mo,W) and X(S,Se) atom as given by Eq. (\ref{eq:SRcore}).  the fitted parameters for all atoms are listed in Table~\ref{tab:table1}:

\begin{table*}[ht]
\caption{\label{tab:table1}Short-Range Part of Core Potentials Fitted result with DFT.}
\begin{ruledtabular}
\begin{tabular*}{\textwidth}{@{\extracolsep{\fill}}lccccccccc}
\hline
Atoms & \( r_{cut} \) (\AA) & Exponent (\(\alpha\)) & \multicolumn{6}{c}{Coefficients} \\
\cline{4-9}
    &   &  & \( C_0 \) & \( C_1 \) & \( C_2 \) & \( C_3 \) & \( C_4 \) & \( C_5 \) \\
\hline
W  & 0.6117 & - & -0.16956 & 2.46980 & -12.20674 & 26.45565 & -22.06776 & - \\
   & 3.0    & 0.8504 & 3.28276  & -5.47670 & 1.71172   & -0.20268 & 0.00824   & - \\
\hline
Mo & 0.7407 & - & -14.64721 & 20.11884 & -53.35016 & 194.88336 & -337.87171 & 212.53631 \\
   & 3.0    & - & -93.30706 & 111.27523 & -109.04823 & 42.67719 & -7.80775 & 2.65095 \\
\hline
Se & 0.9986 & - & -3.40847 & -4.38717 & 15.23014 & -10.28998 & 2.27067 & - \\
   & 1.8892 & - & -3.92653 & -1.61556 & 10.40251 & -7.08323 & 1.78130 & -0.15520 \\
   & 3.0    & - & -1.24505 & 0.84698 & -0.23304 & 0.03224 & -0.00223 & 0.00006 \\
\hline
S  & 0.98996 & - & -10.97583 & -4.21172 & 34.9252 & -31.40929 & 8.56341 & 0.40383 \\
   & 1.49543 & - & -10.97583 & -4.21172 & 34.9252 & -31.40929 & 8.56341 & 0.40383 \\
   & 2.2032  & - & 11.36892  & -17.75451 & 10.48766 & -2.99507 & 0.41763 & -0.02286 \\
\hline
\end{tabular*}
\end{ruledtabular}
\end{table*}
As we have addressed the core components, the remaining terms involve contributions from the Hartree, exchange, and correlation functional, denoted as $\Delta \widetilde{V}_{\text{HXC}}(z,\mathbf{G})$. These are transformed from the real-space effective local potential $V_{\text{L}}(\mathbf{r})$ to reciprocal space using Fourier transform.

In Section II of our methodology, specifically in Eq. (\ref{eq:VHXC_G0}), we discuss the isotropic component where $\mathbf{G} = 0$. This component is effectively described by three Gaussian functions for MoS$_2$, WS$_2$, MoSe$_2$, and WSe$_2$, as detailed in Table~\ref{tab:table2}.

\begin{table*}
\caption{\label{tab:table2}Exponents and Coefficients for MoS$_2$, MoSe$_2$, WS$_2$, and WSe$_2$.}
\begin{ruledtabular}
\begin{tabular}{lccc|ccc}
\multirow{2}{*}{Material} & \multicolumn{3}{c|}{Exponents} & \multicolumn{3}{c}{Coefficients} \\
\cline{2-7}
 & $\alpha_{1}^{M}$ & $\alpha_{2}^{M}$ & $\alpha_{3}^{M}$ & $A_{1}^{M}$ & $A_{2}^{M}$ & $A_{3}^{M}$ \\
\hline
MoS$_2$ & Value & Value & Value & Value & Value & Value \\
MoSe$_2$ & Value & Value & Value & Value & Value & Value \\
WS$_2$ & Value & Value & Value & Value & Value & Value \\
WSe$_2$ & Value & Value & Value & Value & Value & Value \\
\end{tabular}
\end{ruledtabular}
\end{table*}

Now, we seek to quantify the anisotropic term (\(\mathbf{G} \neq 0\)) by subtracting the isotropic part (\(\mathbf{G} = 0\)), as defined in Eq. (\ref{eq:VHXC}), for each TMDC structure under consideration. This comparison is detailed in Eq. (\ref{eq:difVHXCSR}) of Section II.

We have verified that the short-range contribution from X (S, Se) atoms is negligible (error $<$ 0.015 a.u.). Therefore, only contributions from M (W, Mo) atoms are considered for the short-range part. Specifically, for (\(\mathbf{G} \geq 3.1 \text{ a.u.}\)), this short-range component is adequately described by a single Gaussian function centered at \(z = 0\).

The fitted parameters for each structure are presented in Table~\ref{tab:table3}. 

\begin{table*}[ht]
\caption{\label{tab:table3}For universal Short-Range part Exponents and Coefficients for MoS$_2$, MoSe$_2$, WS$_2$, and WSe$_2$.}
\begin{ruledtabular}
\begin{tabular}{lccc}
Material & \multicolumn{2}{c}{Exponents (Å)} & Coefficients \\
\cmidrule(lr){2-3} \cmidrule(lr){4-4}
          & $\alpha_{1}^{M}$ & $\alpha_{2}^{M}$ & $A_{1}^{M}$ \\
\midrule
MoS$_2$  & -- & -- & -- \\
MoSe$_2$ & -- & -- & -- \\
WS$_2$   & -- & -- & -- \\
WSe$_2$  & -- & -- & -- \\
\end{tabular}
\end{ruledtabular}
\end{table*}

After subtracting the short-range potential given by Eq. (\ref{eq:difVHXCSR}) from Eq. (\ref{eq:VHXC}), we obtain the remaining long-range part, denoted as \(\Delta{\widetilde{V}}_{\text{HXC,LR}}\left( z, \mathbf{G} \right)\).
This long-range part includes both the real and imaginary components of the potentials. It is fitted using a polynomial multiplied by a Gaussian function for each shell up to \(\mathbf{G} \leq 3.1 \text{ a.u.}\), corresponding to the 2nd to 5th shells in all crystal structures.
For M atoms (Mo, W), the structure factor contributes only real potentials, as their structure factors are purely real. In contrast, X atoms (S, Se) have both real and imaginary parts in their structure factors.
Our approach involves fitting the difference potentials separately for the imaginary parts of each shell. These are then divided by the phase factor of X atoms and multiplied by the phase factor of M atoms. Similarly, we fit the amplitude for the imaginary parts of the potentials, from which we derive the real parts contributed by X atoms.
Subsequently, we subtract the real part of each shell of \(\Delta{\widetilde{V}}_{\text{HXC,LR}}\left( z, \mathbf{G} \right)\) from the real part contributed by X atoms. This yields the pure contribution of M atoms for each shell in the real part, which is fitted using polynomial multiplied by Gaussian functions.
The fitted parameters for each structure are summarized in Table~\ref{tab:table4} and Table~\ref{tab:table5}.
 
\begin{table*}[ht]
\caption{\label{tab:table4}Long-Range potential contributions with parameters for Exponents and Coefficients for MoS$_2$ and WS$_2$.}
\begin{ruledtabular}
\begin{tabular}{@{}lllcccc@{}}
\toprule
Material & \( \mathbf{G}_{s} \)  & Component & \multicolumn{2}{c}{Exponents} & \multicolumn{2}{c}{Coefficients} \\
\cmidrule(lr){4-5} \cmidrule(lr){6-7}
         &      &           & \( \alpha^{M} \) & \( \alpha^{X} \) & \( A^{M} \) & \( A^{X} \) \\
\midrule
\multirow{12}{*}{MoS$_2$} & 2nd  & S  & ... & ... & ... & ... \\
                          &      &    & ... & ... & ... & ... \\
                          &      &    & ... & ... & ... & ... \\
\cmidrule{2-7}
                          & 2nd  & Mo  & ... & ... & ... & ... \\
                          &      &    & ... & ... & ... & ... \\
                          &      &    & ... & ... & ... & ... \\
\cmidrule{2-7}
                          & 3rd  & Mo  & ... & ... & ... & ... \\
                          &      &    & ... & ... & ... & ... \\
                          &      &    & ... & ... & ... & ... \\
\cmidrule{2-7}
                          & 4th  & S & ... & ... & ... & ... \\
                           &      &    & ... & ... & ... & ... \\
                           &      &    & ... & ... & ... & ... \\
\cmidrule{2-7}
                           & 4th  & Mo & ... & ... & ... & ... \\
                           &      &    & ... & ... & ... & ... \\
                           &      &    & ... & ... & ... & ... \\
\cmidrule{2-7}
                           & 5th  & S & ... & ... & ... & ... \\
                           &      &    & ... & ... & ... & ... \\
                           &      &    & ... & ... & ... & ... \\
\cmidrule{2-7}
                           & 5th  & Mo & ... & ... & ... & ... \\
                           &      &    & ... & ... & ... & ... \\
                           &      &    & ... & ... & ... & ... \\
\midrule
\multirow{12}{*}{WS$_2$} & 2nd  & S  & ... & ... & ... & ... \\
                         &      &    & ... & ... & ... & ... \\
                         &      &    & ... & ... & ... & ... \\
\cmidrule{2-7}
                         & 2nd  & W & ... & ... & ... & ... \\
                         &      &    & ... & ... & ... & ... \\
                         &      &    & ... & ... & ... & ... \\
\cmidrule{2-7}
                         & 3rd  & W  & ... & ... & ... & ... \\
                         &      &    & ... & ... & ... & ... \\
                         &      &    & ... & ... & ... & ... \\
\cmidrule{2-7}
                         & 4th  & S  & ... & ... & ... & ... \\
                         &      &    & ... & ... & ... & ... \\
                         &      &    & ... & ... & ... & ... \\
\cmidrule{2-7}
                         & 4th  & W & ... & ... & ... & ... \\
                         &      &    & ... & ... & ... & ... \\
                         &      &    & ... & ... & ... & ... \\
\cmidrule{2-7}
                         & 5th  & S & ... & ... & ... & ... \\
                         &      &    & ... & ... & ... & ... \\
                         &      &    & ... & ... & ... & ... \\
\bottomrule
\end{tabular}
\end{ruledtabular}
\end{table*}

\begin{table*}[ht]
\caption{\label{tab:table5}Long-Range potential contributions with parameters for Exponents and Coefficients for MoSe$_2$ and WSe$_2$.}
\begin{ruledtabular}
\begin{tabular}{@{}lllcccc@{}}
\toprule
Material & \( \mathbf{G}_{s} \)  & Component & \multicolumn{2}{c}{Exponents} & \multicolumn{2}{c}{Coefficients} \\
\cmidrule(lr){4-5} \cmidrule(lr){6-7}
         &      &           & \( \alpha^{M} \) & \( \alpha^{X} \) & \( A^{M} \) & \( A^{X} \) \\
\midrule
\multirow{12}{*}{MoSe$_2$} & 2nd  & Se  & ... & ... & ... & ... \\
                          &      &    & ... & ... & ... & ... \\
                          &      &    & ... & ... & ... & ... \\
\cmidrule{2-7}
                          & 2nd  & Mo  & ... & ... & ... & ... \\
                          &      &    & ... & ... & ... & ... \\
                          &      &    & ... & ... & ... & ... \\
\cmidrule{2-7}
                          & 3rd  & Mo  & ... & ... & ... & ... \\
                          &      &    & ... & ... & ... & ... \\
                          &      &    & ... & ... & ... & ... \\
\cmidrule{2-7}
                          & 4th  & Se & ... & ... & ... & ... \\
                           &      &    & ... & ... & ... & ... \\
                           &      &    & ... & ... & ... & ... \\
\cmidrule{2-7}
                           & 4th  & Mo & ... & ... & ... & ... \\
                           &      &    & ... & ... & ... & ... \\
                           &      &    & ... & ... & ... & ... \\
\cmidrule{2-7}
                           & 5th  & Se & ... & ... & ... & ... \\
                           &      &    & ... & ... & ... & ... \\
                           &      &    & ... & ... & ... & ... \\
\cmidrule{2-7}
                           & 5th  & Mo & ... & ... & ... & ... \\
                           &      &    & ... & ... & ... & ... \\
                           &      &    & ... & ... & ... & ... \\
\midrule
\multirow{12}{*}{WSe$_2$} & 2nd  & Se  & ... & ... & ... & ... \\
                         &      &    & ... & ... & ... & ... \\
                         &      &    & ... & ... & ... & ... \\
\cmidrule{2-7}
                         & 2nd  & W & ... & ... & ... & ... \\
                         &      &    & ... & ... & ... & ... \\
                         &      &    & ... & ... & ... & ... \\
\cmidrule{2-7}
                         & 3rd  & W  & ... & ... & ... & ... \\
                         &      &    & ... & ... & ... & ... \\
                         &      &    & ... & ... & ... & ... \\
\cmidrule{2-7}
                         & 4th  & Se  & ... & ... & ... & ... \\
                         &      &    & ... & ... & ... & ... \\
                         &      &    & ... & ... & ... & ... \\
\cmidrule{2-7}
                         & 4th  & W & ... & ... & ... & ... \\
                         &      &    & ... & ... & ... & ... \\
                         &      &    & ... & ... & ... & ... \\
\cmidrule{2-7}
                         & 5th  & Se & ... & ... & ... & ... \\
                         &      &    & ... & ... & ... & ... \\
                         &      &    & ... & ... & ... & ... \\
\bottomrule
\end{tabular}
\end{ruledtabular}
\end{table*}

For the nonlocal part of the scalar-relativistic pseudopotential defined in Sec II of Eq.~(\ref{eq:nonlocalpot}), we need to fit the \(\left| \beta_{lm}^{n\sigma} \right\rangle\) projector for each angular momentum \(l\), in order to obtain the contribution due to core part NLPPs. The fitted parameters, in the form of Eqs.~(\ref{eq:nonlocalseg1}) and (\ref{eq:nonlocalseg2}), are provided in the supplementary information.

\subsubsection{\textbf{BAND STRUCTURES}}
The all-electron (AE) Bloch states of graphene are written as
\begin{equation}
\begin{split}
\psi_{\nu,\mathbf{k}}\left( \mathbf{\rho},z \right) &= (1 + \widehat{S})\varphi_{\nu,\mathbf{k}}\left( \mathbf{\rho},z \right) \\
&= (1 + \widehat{S})\sum_{i,\mathbf{G}}^{}Z_{i\mathbf{G}}^{\nu,\ \mathbf{k}}B_{i}(z)\frac{1}{\sqrt{A}}e^{i\left( \mathbf{k} + \mathbf{G} \right) \cdot \mathbf{\rho}},
\end{split}
\end{equation}
where \(\varphi_{\nu,\mathbf{k}}\left( \mathbf{\rho},z \right)\) denotes
the pseudo-wavefunction, \(B_{i}(z)\) denotes the B-spline function and
\(\widehat{S}\ \)is the projection operator defined in (?). The
all-electron Schr\(\ddot{o}\)dinger equation reads
\begin{equation}
{\widehat{H}}_{AE}\psi_{\nu,\mathbf{k}}\left( \mathbf{\rho},z \right) = E_{\nu}(\mathbf{k})\psi_{\nu,\mathbf{k}}\left( \mathbf{\rho},z \right)
\end{equation}
\(where\ {\widehat{H}}_{AE}\) is all-electron Hamiltonian operator and
\(E_{\nu}(\mathbf{k})\) denotes the energy of the \( \nu \)-th band at wavevector \( \mathbf{k} \). Substituting .. gives rise to the following generalized eigenvalue problem
\begin{equation}
\begin{split}
& \sum_{i',\mathbf{G}'}^{}\left\langle \mathbf{k} + \mathbf{G};B_{i} \middle| \widehat{H} \middle| \mathbf{k} + \mathbf{G}';B_{i'} \right\rangle Z_{i'\mathbf{G'}}^{\nu,\ \mathbf{k}} \\
& \quad = E_{\nu}\left( \mathbf{k} \right)\sum_{i^{'}\mathbf{G}^{'}}^{}\left\langle \mathbf{k} + \mathbf{G};B_{i} \middle| 1 + \widehat{S} \middle| \mathbf{k} + \mathbf{G}^{'};B_{i'} \right\rangle Z_{i^{'}\mathbf{G}^{\mathbf{'}}}^{\nu,\ \mathbf{k}}
\end{split}
\end{equation}
where
\(\widehat{H} = \widehat{T} + {\widehat{V}}_{loc} + {\widehat{V}}_{nl}\)
is the pseudo Hamiltonian. \(E_{\nu}\left( \mathbf{k} \right)\) denote
the eigenvalues and \(Z_{i'\mathbf{G'}}^{\nu,\ \mathbf{k}}\) the
eigenvectors.

We first solve the generalized eigenvalue problem in the scalar relativitic (SR) approximation in which the spin-orbit Coupling (SOC) is neglected. To reduce computation time, we take symmetric and
antisymmetric combinations of localized B-spline basis functions related by the mirror operation with respect to z axis to form symmetrized B-splines. Namely,
\begin{equation}
\begin{split}
B_{i}^{\pm}(z) &= \left[ B_{i}(z) \pm B_{\overline{i}}(z) \right]/\sqrt{2} \\
& \text{for } i = 1, (N - 1)/2 \text{ and } \overline{i} = N + 1 - i.
\end{split}
\end{equation}
where \(N\) is the number of B-splines adopted. Due to the mirror
symmetry of monolayer TMDCs, the eigenstates of \(\widehat{H}\) can be grouped into even and odd states and the two sets of states are decoupled. Within the symmetrized B-spline basis, the computation time can be reduced by a factor four.

To include the effect of SOC, we expand the eigen-spinors of the fully relativistic Hamiltonian in term of linear combinations of the eigenfunctions obtain in the scalar relativistic approximation as follows
\begin{equation}
\Psi_{n,\mathbf{k}}\left( \mathbf{\rho},z \right) = \sum_{\nu}^{}
\begin{pmatrix}
C_{\nu,\mathbf{k}}^{+} \\
C_{\nu,\mathbf{k}}^{-} \\
\end{pmatrix}
\psi_{\nu,\mathbf{k}}\left( \mathbf{\rho},z \right)
\end{equation}
where \(C_{\nu,\mathbf{k}}^{+}\) and \(C_{\nu,\mathbf{k}}^{-}\) denote
the expansion coefficients for the spin-up and spin-down component respectively. Substituting Eq.   into the all-electron Schr\(\ddot{o}\)dinger equation gives
\begin{equation}
\begin{split}
& \sum_{\nu}^{}\left\langle \varphi_{\nu',\mathbf{k}} \middle| H_{2 \times 2} \middle| \varphi_{\nu,\mathbf{k}} \right\rangle
\begin{pmatrix}
C_{\nu,\mathbf{k}}^{+} \\
C_{\nu,\mathbf{k}}^{-} \\
\end{pmatrix} \\
&= {\widetilde{E}}_{n}\left( \mathbf{k} \right)
\sum_{\nu}^{}\left\langle \varphi_{\nu',\mathbf{k}} \middle| I_{2 \times 2} + S_{2 \times 2} \middle| \varphi_{\nu,\mathbf{k}} \right\rangle
\begin{pmatrix}
C_{\nu,\mathbf{k}}^{+} \\
C_{\nu,\mathbf{k}}^{-} \\
\end{pmatrix}
\end{split}
\end{equation}
where
\(H_{2 \times 2}\)=\({\widehat{H}}_{sR}I_{2 \times 2} + \mathrm{\Delta}V_{2 \times 2}^{nl}\)
denotes the fully relativistic pseudo Hamiltonian, \(I_{2 \times 2}\)
denotes the identity matrix in the \(2 \times 2\) spin space, and
\({\widehat{H}}_{sR}\) denotes the scalar relativistic pseudo
Hamiltonian.
\(\mathrm{\Delta}V_{2 \times 2}^{nl} = V_{2 \times 2}^{nl} - {\widehat{V}}_{nl}I_{2 \times 2}\)
denotes the difference in nolocal pseudopoetntial between the fully
relativistic case and scalar relativistic case. The matrix elements of
\({\widehat{V}}_{nl}\) and \(V_{2 \times 2}^{nl}\) in the mixed basis
\{\(\mathbf{K};B_{i}\}\) have been given in Eq.  and Eq.
respectively.

Substituting the relation
\({\widehat{H}}_{sR}\varphi_{\nu,\mathbf{k}}\left( \mathbf{\rho},z \right) = E_{\nu}\left( \mathbf{k} \right)(1 + \widehat{S})\varphi_{\nu,\mathbf{k}}\left( \mathbf{\rho},z \right)\)
 yields
\begin{equation}
\begin{split}
& \sum_{\nu}^{}\left\lbrack E_{\nu}\left( \mathbf{k} \right)\delta_{\nu',\nu} + \left\langle \varphi_{\nu',\mathbf{k}} \middle| \mathrm{\Delta}V_{2 \times 2}^{nl} \middle| \varphi_{\nu,\mathbf{k}} \right\rangle \right\rbrack  \begin{pmatrix}
C_{\nu,\mathbf{k}}^{+} \\
C_{\nu,\mathbf{k}}^{-} \\
\end{pmatrix} \\
&= {\widetilde{E}}_{n}\left( \mathbf{k} \right)\sum_{\nu}^{}\left\lbrack \delta_{\nu',\nu} + \left\langle \varphi_{\nu',\mathbf{k}} \middle| \mathrm{\Delta}S_{2 \times 2} \middle| \varphi_{\nu,\mathbf{k}} \right\rangle \right\rbrack  \begin{pmatrix}
C_{\nu,\mathbf{k}}^{+} \\
C_{\nu,\mathbf{k}}^{-} \\
\end{pmatrix}
\end{split}
\end{equation}
where
\(\mathrm{\Delta}S_{2 \times 2} = S_{2 \times 2} - \widehat{S}I_{2 \times 2}\)\emph{.}
The matrix elements of \(\widehat{S}\) and \(S_{2 \times 2}\) in the
mixed basis \{\(\mathbf{K};B_{i}\}\) have been given in Equations respectively.

\begin{table*}[ht]
\caption{\label{tab:table6}Structural parameters (\(a\), X-X), spin-orbit coupling (SOC), and band gap (\(E_g\)) values for layer TMDC materials.}
\begin{ruledtabular}
\begin{tabular}{lccccccc}
Material & \multicolumn{2}{c}{Lattice Parameters (Å)} & \multicolumn{2}{c}{SOC (meV)} & \multicolumn{3}{c}{\(E_g\) (eV)} \\
\cmidrule(lr){2-3} \cmidrule(lr){4-5} \cmidrule(lr){6-8}
         & \(a\) & X-X & PBE & SEPM & PBE & SEPM & Experimental \\
\midrule
MoS$_2$  & 3.160 & 3.172 & -- & -- & -- & -- & -- \\
MoSe$_2$ & 3.299 & 3.338 & -- & -- & -- & -- & -- \\
WS$_2$   & 3.153 & 3.166 & -- & -- & -- & -- & -- \\
WSe$_2$  & 3.282 & 3.340 & -- & -- & -- & -- & -- \\
\bottomrule
\end{tabular}
\end{ruledtabular}
\end{table*}
\section{CONCLUSION}
Here,
I have to put these paper citation \cite{shewchuk1994introduction,hamann1979norm,kresse1999ultrasoft}  to get rid of arxiv submission. I will put later appropriately.
\nocite{*}

\providecommand{\noopsort}[1]{}\providecommand{\singleletter}[1]{#1}%

\end{document}